\documentclass{elsart}

\usepackage{natbib}

 \usepackage{epsfig}

\usepackage{amssymb}

\begin{document}

\begin{frontmatter}


\title{Alternative Data Reduction Procedures for UVES: Wavelength 
Calibration and Spectrum Addition}~\footnote{
Based on observations made with ESO Telescopes at the La Silla or
Paranal Observatories under program IDs 68.A-0106 and 70.A-0017}
\author{Rodger I. Thompson}
\address{Steward Observatory, University of Arizona, Tucson, 
Arizona 85721, USA}
\author{Jill Bechtold}
\address{Steward Observatory, University of Arizona, Tucson, 
Arizona 85721, USA}
\author{John H. Black}
\address{Department of Radio and Space Science, Chalmers University
of Technology, Onsala Space Observatory, SE-43992, Sweden}
\author{C.J.A.P. Martins}
\address{Centro de Astrof\'{\i}sica, Universidade do Porto, Rua  
das Estrelas, 4150-762 Porto, Portugal and DAMTP, University of Cambridge, 
Wilberforce Road, Cambridge CB3 0WA, United Kingdom}
\ead{rthompson@as.arizona.edu}
\ead{jbechtold@as.arizona.edu}
\ead{John.Black@chalmers.se}
\ead{C.J.A.P.Martins@damtp.cam.ac.uk}

\begin{abstract}

This paper addresses alternative procedures to the ESO supplied
pipeline procedures for the reduction of UVES spectra of two quasar
spectra to determine the value of the fundamental constant 
$\mu = M_p/M_e$ at early times in the universe.  The procedures 
utilize intermediate product images and spectra produced by the 
pipeline with alternative wavelength calibration and spectrum addition methods.
Spectroscopic studies that require extreme wavelength precision
need customized wavelength calibration procedures beyond that
usually supplied by the standard data reduction pipelines. An
example of such studies is the measurement of the values
of the fundamental constants at early times in the universe. 
This article describes a wavelength calibration procedure for
the UV-Visual Echelle Spectrometer on the Very Large Telescope,
however, it can be extended to other spectrometers as well.
The procedure described here provides relative wavelength 
precision of better than $3\times 10^{-7}$ for the long-slit 
Thorium-Argon calibration lamp exposures.  The gain in precision
over the pipeline wavelength calibration is almost entirely
due to a more exclusive selection of Th/Ar calibration lines.  

\end{abstract}


\end{frontmatter}

\section{Introduction} \label{s-int}

This paper serves two main purposes.  The first is to establish 
a detailed record of the UVES wavelength calibration and spectrum 
addition procedures used in the measurement of the fundamental constant 
$\mu \equiv M_p / M_e$ \citep{thm08} based on spectra taken with UVES on 
the VLT.  The second purpose is a description of an alternative to the 
pipeline method of establishing a wavelength calibration of Very Large 
Telescope (VLT) UV-visual Echelle Spectrometer (UVES) spectra.  In a 
belief that a scientific result on the value of a fundamental constant in
the early universe should be described well enough that it
can be repeated by other researchers, we have provided more
detail than is usually present in the description of a data
reduction method.

The standard data reduction pipeline provided by the European 
Southern Observatory (ESO) for UVES does an excellent job of 
providing well calibrated spectra for use in most scientific 
investigations.  Certain investigations, however, require a more 
precise wavelength calibration than is provided by the pipeline.  
\citet{mur07} point out that the UVES pipeline wavelength calibration
has systematic deviations in accuracy that can adversely affect
precise measurements.  In addition, proper validation of the accuracy of
results obtained from a pipeline data reduction requires an exact 
knowledge of the pipeline data reduction procedures. This is often 
not easily available or discernible
for a standard pipeline.  One example of such observations is the 
measurement of the values of fundamental constants in the early
universe such as $\alpha$, the fine structure constant and $\mu$, the 
ratio of the proton to electron mass, through precise spectroscopy of
distant objects.  The faintness of such objects often makes the use
of techniques such as iodine absorption cells impractical.  This 
article describes a set of observing and data reduction procedures
that can improve upon the wavelength accuracy relative to that
available with the standard UVES pipeline.  The procedures achieve
a wavelength accuracy of $\Delta \lambda / \lambda \approx 2.5 \times 10^{-7}$
for Thorium-Argon lines in the long-slit calibration spectra for
UVES with a slit width of 0.8 arc seconds.  The main component in 
achieving this accuracy is the proper selection of appropriate
calibration lines, similar to that of \citet{mur07}.

The examples used in this analysis are VLT UVES spectra of Q0347-383 and
Q0405-443 taken in January of 2002 and 2003 respectively\footnote{Based on
observations made with ESO telescopes at the Paranal Observatory under
program IDs 68.A-0106 and 70.A-0017}.  These particular spectra have been 
used to determine the value of $\mu$ at early times in the universe 
(\citet{rei06} and references therein, \citet{kin08}).  The measurements of
$\mu$ are based on the absorption line wavelengths of molecular hydrogen
produced in damped Lyman alpha clouds along the line of sight to the 
QSO.  The redshifts of the H$_2$ lines are z=3.0249007 for Q0347-383 and
z=2.5947361 for Q0405-443 \citep{thm08}.  In this work, however, we concentrate
on the long-slit calibration line spectra that accompany these observations
and the proper addition of the individual 2x2 binned pixel object spectra
that are an intermediate product of the UVES pipeline data reduction.  In all
of the following the word pixel refers to the 2x2 binned pixels produced during 
readout and maintained during the pipeline processing.
The goal is to find, as accurately as possible, the true wavelength solution
for each order of the spectrum and how that solution varies with time during
the course of the observations.  All solutions are for vacuum wavelengths.
The wavelength calibration techniques are independent of the
objects being observed. The best results occur when a rigorous sequence of
object and calibration observations is performed.  This sequence was not
strictly followed in this case but the sequence is close enough to
be useful.

\section{The observations} \label{s-obs}

The observations of Q0347-383 and Q0405-443 with UVES on VLT occurred during the
nights of January 7-9 2002 for Q0347-383 and January 4-6 2003 for Q0405-443
from a program described by \citet{pet04} and \citet{iva05}.  On each of 
the nights three separate spectra of the QSO were taken with accompanying 
short and long-slit calibration lamp integrations at the same grating setting,
all with a slit width of 0.8 arc seconds. 
We note that the long-slit calibration lamp spectra are included in the observation
definition, presumably to prevent a resetting of the grating position between
the object and calibration spectra.  In this analysis we only use the long
slit calibration spectra which have a higher signal-to-noise ratio than the short
slit spectra. For Q0347-383 two long-slit calibration spectra were taken each
night immediately after the first and second object spectra with the third object
spectrum occurring immediately after the second long-slit calibration spectrum.
For Q0405-443 three long-slit calibration spectra were taken on the first and 
third night immediately after each of the three object spectra.  On the second
night there is no long-slit calibration spectrum after the second object spectrum
so only two long-slit spectra exist for that night. This appears to contradict
the statement in \citet{iva05} which states that calibration spectra occurred
before and after each object spectrum but it is possible that there are other 
calibration images that the archive did not associate with the QSO integrations.
The Q0347-383 observations were taken at a grating setting that put the center
wavelength at 4300 \AA\ and the Q0405-443 observations at a center wavelength
of 3900 \AA. We will refer to these as the 430 and 390 grating settings in the
following.  We use only the blue channel spectra since they contain the 
spectral area of interest for the measurement of $\mu$.  Table~\ref{tab-lag}
gives some of the observing parameters for the long-slit calibration lamp
images.

\subsection{Suggested Observing Procedure for Calibration Spectra} \label{ss-obp}

For very accurate wavelength calibration we suggest an observing sequence that
takes several (more than five) long-slit calibration line spectra immediately
before and immediately after each spectral image of the observed object.  The
primary purpose of the multiple calibration line images is to 
produce a meaningful median image that is free of cosmic ray events
rather than an increase in the signal to noise ratio of the image.  The 
time pacing of the calibration images also tracks drifts in the wavelength 
calibration due to any contributing factor.  The small amount of extra time 
devoted to this observing sequence can make the production of cosmic ray free
images straight forward. Observing the calibration lamp immediately before and
immediately after an object observation provides an accurate monitor 
of any time variation in the wavelength calibration. If the object observations
are spaced close together one set of calibration lamp exposures can serve as
both the before and after function for intermediate object exposures.

\section{The Pipeline Data reduction} \label{s-dr}

All files associated with the observations of the two QSOs were downloaded from
the ESO data archive. The MIDAS based UVES pipeline reduction software was
also downloaded and installed in a LINUX based operating system.
All of the data were processed with the pipeline software which removes the bias
and dark counts.   The bias and dark corrected files are designated by the the 
suffix b\_.bdf in their file names.  All of the image files are converted to 
FITS format files using the MIDAS OUTDISK/FITS procedure.  Our analysis
starts with the two dimensional FITS format images of the 
long-slit calibration lamp images and long-slit flat-field images that have 
been bias and dark subtracted but not flat fielded.  These files are identified 
by the terms LAMP,WAVE and LAMP,FLAT in the OBJECT parameter in the file header. 
We also utilize the short slit lamp flats identified by LAMP,ORDERDEF for 
identifying the location of the orders. A pipeline produced table of x and y
line positions is used to provide a first guess and the wavelength solution.
In this and in all other references to x and y, the x direction is along the
dispersion direction and the y direction is perpendicular to the dispersion
direction as seen in the pipeline rotated images.  The order
traces are generally along the x direction but are curved and tilted upward.
The pipeline produced intermediate product spectra are used to
produce the final spectrum.  We do not use the primary output of the pipeline
which are spectra interpolated to an equally spaced wavelength grid.

\section{Wavelength Calibration} \label{s-wc}

\begin{figure}
\includegraphics[scale=.6]{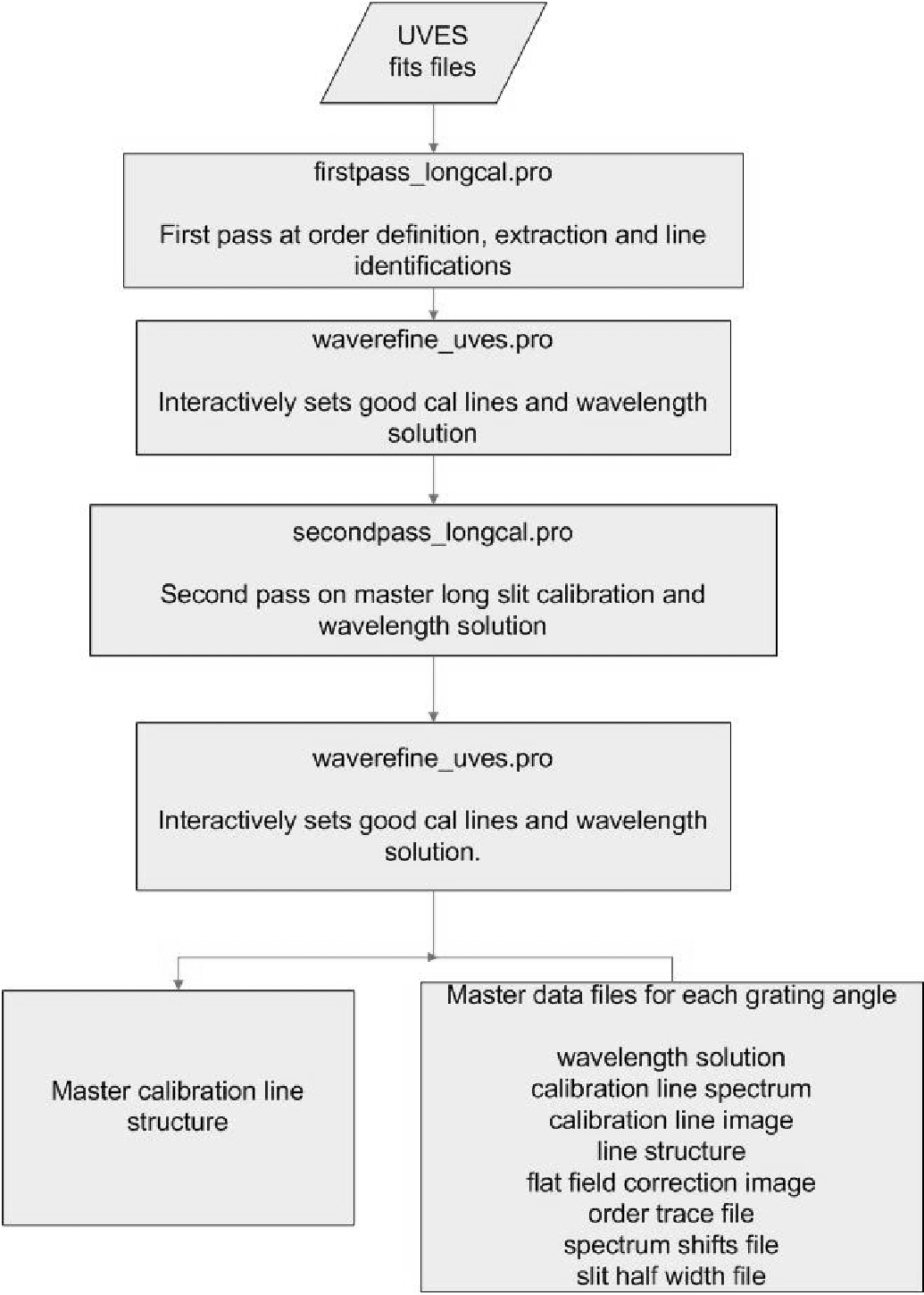}
\caption{Overview flow chart for the wavelength calibration process.
The names that end in .pro at the top of the boxes are the names of
the IDL based procedures that were developed by the authors.}
\label{fig-ov}
\end{figure}

All procedures are written in IDL\footnote{IDL stands for Interactive Data 
Language registered by Research Systems Inc.} code. In the following we
will write IDL provided procedure names in capital italic letters and 
procedures written by the authors using IDL code in lower case italics.
After further refinement the authors intend to make the reduction code
available to the public, hence the use of the procedure names in the
flow charts.
Following the production of the pipeline products the wavelength calibration
proceeds in two semi-independent phases.  The first phase, called the 
first pass, uses the long-slit bias corrected images 
produced by the pipeline reduction.    In the first pass the
long-slit calibration images for a given setting are median combined to
form a ``master'' long-slit calibration image that is free of cosmic ray
hits.  A first guess at the wavelength solution based on the line positions
output by the pipeline provides the initial wavelength solution
for each order. The solution is then iterated interactively
and a set of ``good'' calibration lines is identified.  This step involves
some subjective judgment on the definition of good that is discussed below.
In the second phase, called the second pass, the exact spatial locations 
of the good lines are determined. Each long-slit calibration image is 
shifted to a common position and a new median image is produced. The shifts 
are typically on the order of a few hundredths of a pixel. The wavelength 
solution is again interactively iterated to provide the final solution 
for each order. The primary output is a high signal-to-noise ratio master 
long-slit calibration lamp image and a vacuum wavelength solution
for each order. The method assumes that the differences between two 
calibration lamp images at the same grating setting are small in order to
form a useful median image at the beginning of the first pass data reduction.
Once the master long-slit calibration line image has been produced, however,
the actual shifts of subsequent images can be accurately determined from the
master. An overview flow chart of this process is shown in
Figure~\ref{fig-ov}.  The individual procedures of the analysis are described
in subsequent sections. 

\subsection{First pass wavelength solution} \label{s-fp}

\begin{figure}
\includegraphics[scale=.6]{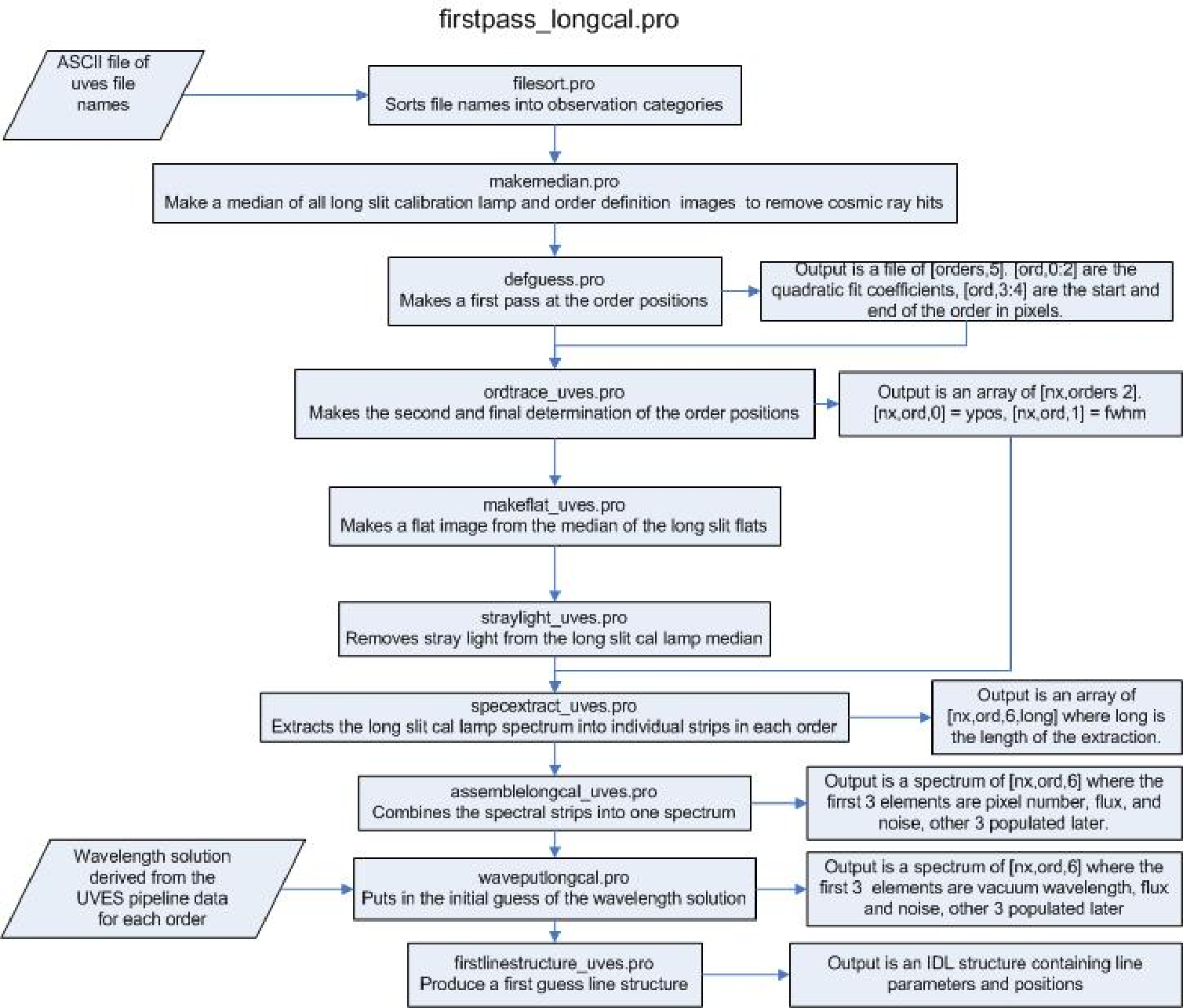}
\caption{Flow chart for the first pass analysis of the long-slit calibration
lamp spectrum and wavelength calibration. The names that end in .pro at the
top of the boxes are the names of the IDL based procedures that we developed.}
\label{fig-fp}
\end{figure}

The first phase of the method, after producing the pipeline products, is to 
make a first pass at the wavelength solution which will be improved in
the second pass.  This step can be eliminated in subsequent measurements at 
the same grating setting once a master long-slit calibration line image and 
wavelength solution have been established. A flow diagram of the first pass
wavelength calibration procedures is shown in Figure~\ref{fig-fp}.  The names
in the boxes that end in .pro are simply the names of the IDL based procedures
that were developed for this process. 

The first pass starts by sorting
all of the two dimensional images to determine their type by
the entry into the OBJECT parameter of the image headers.  All of the long
slit calibration lamp images are median combined to eliminate the 
cosmic ray events.  There
is no attempt in this first pass to adjust the images for small shifts in
the location of the lines.  The positions of the lines typically had shifts
of a few hundredths of a pixel over a three-night set of images.  The 390 grating
setting median contains 6 images and the 430 setting image contains 8 images.
Visual inspection of the images did not detect any residual cosmic ray hits.

\subsubsection{Order definition} \label{ss-od}

\begin{figure}
\includegraphics[scale=.8]{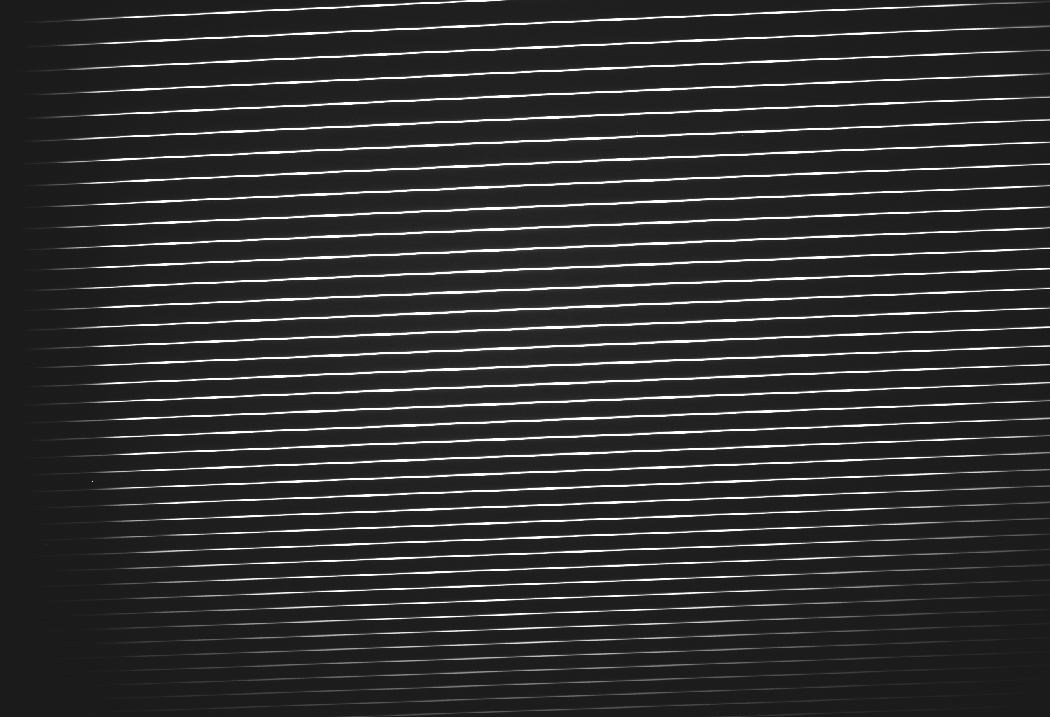}
\caption{The order definition image for UVES at the 390 grating setting 
(centered at 390 nm).  Note the variation in intensity in both the horizontal 
and vertical directions and the slight upward tilt of the orders.}
\label{fig-ord}
\end{figure}

The location of each of the spectral orders must be determined to start the 
spectrum fitting. The short slit flat lamp images are used to determine the
order positions.  The first step is to median the images together to
elliminate cosmic ray events.  One 
order definition image was taken in each night therefore the medians 
contain three images which is a minimum, but certainly not optimal, number 
of images to form a median.  Figure~\ref{fig-ord} shows the $1500 \times 
1024$ binned pixel 
median order-definition-image for the 390 grating setting. Figure~\ref{fig-ord} 
shows that the intensity of the order definition image varies strongly over the
image in both the horizontal and vertical directions.  This is a challenge
to an automated order finding algorithm.  The intensity variation of the 
spectra is similar and results in the first 100 pixels of the spectrum being
rejected due to a poor signal-to-noise ratio.  The orders are curved and are slanted
upward going from left to right.

Next a first solution to the position of the orders is found.  At this point 
the procedure becomes partially UVES specific.  The procedure knows that the 
pipeline has rotated the images so that the orders run horizontally from left 
to right with a slight upward tilt.  It therefore looks for new orders appearing 
from the bottom of the image and knows that it should look for orders along 
the vertical 
direction. An efficiency map is formed by smoothing the order image with a 101 by
101 box four times.  The search for orders starts 100 pixels past the left
edge of the order image.  The region past 100 pixels is binned in groups of
10 columns each and the groups are summed to form single columns of higher signal
to noise for both the original order image and the smoothed image.  The derivative
is taken of each order column and the order position is defined by the
S-shaped curve produced when an order is encountered.  The order position is 
where the value of the derivative changes sign between positive and negative.  
The derivative 
must have at least two positive values followed by two negative values to be
considered an order position.  The flux at that position in the summed order 
column is compared to the flux in the smoothed column to check against
noise that could produce the same pattern of two positive values followed by
two negative values in the derivative.  If an order is found below the lowest
existing order a new order is declared and it becomes the lowest existing order.

After all of the order points have been found the orders are fit in x and y with
a quadratic fit.  The fit is displayed on the computer screen overlaying the
order image to check for any bad fits.  The output is an array  which has the
quadratic coefficients plus the pixel numbers of the starting and ending 
pixels where the order is valid.
Orders that run off the top of the image have ending pixel values less than the
1500 pixel extent of the image.  This array is used as the starting 
point of the final step in the order definition.

After the initial order definition step the y position of each order for each
pixel column is defined.  At each order position defined by the quadratic fit a
Gaussian fit is found in an 11 pixel column centered on the pixel position given
by the fit.  The IDL procedure \emph{GAUSSFIT} fits a Gaussian plus a constant offset to
the column to determine the y position of the peak.  After the x and y position 
for each order is determined by Gaussian fitting each order is fit with a
six term Legendre polynomial fit using the IDL procedure \emph{SVDFIT} with the Legendre
option set.  The Legendre polynomial order fit gives the y position in pixels 
of each order for every pixel column.  This is then taken as the vertical 
center position of the order at each horizontal pixel position in the long 
slit calibration lamp images in the following analysis.

\subsubsection{Flat field} \label{ss-ff}

Although not absolutely critical to the wavelength calibration, the next step in
the analysis is the production of the flat-field.  This is identical to the flat
field used in the object spectrum analysis, therefore it is also used in extracting
the calibration lamp spectrum for consistency.  The flat-field is constructed from
the median of all of the long-slit flat-field images taken at the desired grating
setting.  The procedure first determines the length of the slit by looking at the
center of the middle order.  It determines along a vertical line where the intensity
of the image falls to 1/2 of the maximum and defines the height between those points
as the length of the slit.  The procedure produces a flat correction image which
is the inverse of the long-slit flat lamp image in the region of each order defined
by the length of the slit and the beginning and ending of the order as defined in
\S~\ref{ss-od}.  The image is divided by the median of all of the non-zero
pixels in the flat correction image to make the median of the non-zero pixels
equal to one.  Correction of the object spectra is achieved by multiplying the
object spectrum image by the flat correction image.

\subsubsection{Stray light} \label{ss-sl}

The long-slit calibration lamp images have low level background stray light from
nonspecular reflections, grating imperfections and other sources.  The next step
in the analysis removes the smooth component of the stray background light to first
order.  The photon noise from the background light is, of course, still in the 
stray-light-corrected image.  To start the procedure the median long-slit
calibration lamp image is multiplied by a mask that is zero in the regions that
contain the calibration lamp spectra and 1 in the inter-order regions.  We assume
that the stray light observed in the inter-order region extends smoothly into
the regions containing the orders.  The masked image is divided into regions
of 10 columns each.  Each region of 10 columns is median combined in the 
horizontal direction to form single columns of higher signal-to-noise ratio.  A third
order polynomial is fit along the vertical direction of each of the
columns using only the non-zero inter-order regions of the columns.  This produces
a set of measures of the stray light in the vertical direction spaced every 10
columns in the middle of each 10 column region.  This results in a stray light
image that has more resolution in the horizontal direction than in the vertical.
This was done because the variation of the stray light appears to be stronger
in the horizontal direction than in the vertical. The final stray light image is
formed by interpolating the solution in the horizontal direction for each
row of the image.  This smooth distribution is then subtracted from the median
long-slit calibration lamp image.

\subsubsection{Extraction of the spectrum}

The spectrum extraction procedure produces a two dimensional image of each order.
The height of the image is the slit length in pixels and its length is the length
of the spectral image in pixels.  The total output is a four dimensional array
with the dimensions spec[nx,norders,6,length] where nx is the number of pixels in the
x direction, norders is the number of usable orders and length is the length of the
slit in pixels. For the 6 inputs in the third index the first is the wavelength,
which at this stage is just the pixel number, the second is the flux, the third
the noise, and the other 3 are populated later in other spectral analysis procedures.
The first step in the procedure is to multiply the total spectral image by the 
flat-field correction image described in \S~\ref{ss-ff}. 

The procedure next extracts the two dimension spectral image of each order. The center
pixel in the vertical direction for each order is determined by the order solution
discussed in \S~\ref{ss-od}.  The IDL procedure ROUND finds the pixel number and
the pixels above and below the center pixel are extracted according the the value
of the slit length in pixels.  The flux is in ADUs (Analog to Digital converter
Units).  The spectral image of the fifth order in the array is
spec[0:nx-1,4,1,0:length-1] in the IDL array. In IDL the nomenclature a:b
means all entries with indexes a through b.  The first index in IDL is 0 rather
than 1 which accounts for the length-1 in the preceding and why the fifth order has 
index 4.  The noise for each pixel is computed as the root mean square of
the photon and read noise.  The flux in ADUs per second is converted to electrons
using the conversion factor in the header and the flat-field correction.  The read 
noise is also listed in the header.  The calculated noise in electrons is the
square root of the sum of the total number of electrons plus the square of the
read noise.  The noise in electrons is converted back into ADUs and re-corrected
for the flat-field.  The noise image is in the same format as the flux
except that the third index in the extraction array is 2 rather than 1.

\subsubsection{Spectrum assembly}

The next step is the creation of a one dimensional spectrum for each order from
the individual strips of spectra created by the extraction.  Inspection of the
long-slit calibration line image indicates that the intensity along the
length of the slit is quite uniform and that the image of the slit is vertical
with no apparent deviation over the entire field of view.  For these reasons
the assembled one dimensional spectrum is simply the sum of all of the pixels
in the vertical direction.  If there is any fractional pixel tilt in any 
part of the field of view it would result in a slightly lower resolution but
the center point of the line would still be the same.  A comparison of the
object spectra with the order definition positions confirmed that the object
spectra maxima lie at the middle of the slit length. The noise is
simply the square root of the sum of the squares of the noise determined
for each pixel during the spectrum extraction.  After this stage we have
individual one dimensional spectra for each order.  The physical orders
are 141 to 102 for the 390 grating setting and 127 to 95 for the 430 grating
setting, however only a subset of the orders had object spectra appropriate
for the determination of the value of $\mu$ at the redshifts of the absorption
line systems.  We list the order numbers in inverse order since the 
image orientation returned by the pipeline has the order number decreasing 
in the vertical direction. The spectrum data cube now has the dimensions 
spec[nx,nord,6] since the previous array was summed along its fourth 
dimension.  Note that nord equal to zero corresponds to order 141 and 127 for 
the 390 and 430 grating settings. A value of nord equal to 1 corresponds to orders
140 and 126 respectively etc. 

\subsubsection{Preliminary wavelength assignment}

The UVES pipeline reduction produces a table of the x and y positions
for calibration lines during the pipeline reduction along with the 
air wavelengths of the line and the order number.  The table is 
labeled lxxxBLUE.tbl for the blue channel where xxx is the grating 
setting such as 390.  We use the procedure \emph{lineguess.pro} on
the output of this table to produce a preliminary wavelength solution
for each order.  The first step is the conversion of the air wavelengths
to vacuum wavelengths by the formulas of \citet{edl66} as given in the
\citet{crc}.  The \emph{lineguess} procedure uses the IDL 
procedure \emph{SVDFIT.PRO} 
in double precision with the Legendre polynomial option set to produce
a six term Legendre polynomial fit to the wavelengths.  The outputs
of the procedure are six coefficients of the first six Legendre polynomials
for each order.  These coefficients are the inputs to the procedure
\emph{waveput1m1.pro} which converts the pixel numbers in the wavelength
array of the spectrum to wavelengths. The wavelength for each of
the nx pixels for an order nord in the output data cube is given by
spec[0:nx-1,nord-1,0].  See \S~\ref{ss-rw} for more details of the 
wavelength fitting.

\subsubsection{Thorium and Argon line lists} \label{sss-ta}

An important part of the wavelength calibration improvement over
the pipeline calibration is the use of newly published line lists.
The Thorium and Argon line lists came from two sources, \citet{lov07}
and \citet{mur07}. \citet{lov07} list all of the Thorium-Argon lines
between 3785.6 and 6914.4 \AA\ found in a dedicated observation of 
Thorium-Argon lines with the HARP spectrometer.  The line list from
\cite{mur07} lists lines between 3033.5 and 10506.5 \AA. This list
only includes lines that have been deemed acceptable in \cite{mur07}
for accurate wavelength calibration.  Our lists were constructed from
files kindly sent by Michael Murphy prior to publication, however, the
files are now available from websites listed in the publications.  Our
final list of lines contains all of the lines in the \citet{lov07} list
plus the lines from \citet{mur07} that lie at shorter wavelengths than
the \citet{lov07} list.  The data from the lists for each line are entered 
into an IDL data structure.   An IDL data structure is a list of parameters 
that can be a mixed combination of numbers, text strings or arrays, each 
with a different format. The structures for all of the identified lines are 
identical in format and the entire set of lines forms an IDL structure array 
for easy access by the IDL based analysis procedures. The data items for each 
line in the IDL structure are: the 
wavenumber in cm$^{-1}$; the vacuum wavelength in \AA, the standard pressure
and temperature (STP) air wavelength, 
the previously determined wavelength listed by \citet{lov07}, the error
from \citet{lov07}, the intensity, the identification, and a ``Murphy'' 
flag.  The ``Murphy'' flag is the letter M if it is listed by \citet{mur07} 
as a good line and the letter N if it is not.  Note that the \citet{lov07}
list contains all of the observed lines in its wavelength range while \citet{mur07}
only list the good lines therefore all of the lines outside the range of the 
\citet{lov07} lists are ``Murphy'' lines. 

\subsubsection{First determination of the line positions and identifications}

The next step in the first pass through the wavelength calibration is
the location of the calibration line positions in the median long-slit
calibration line image and a preliminary identification of the proper 
Thorium or Argon line for the line position from the line list 
described in \S~\ref{sss-ta}.  

The line identification procedure for a particular order first determines
the median flux in the order which is assumed to be the continuum emission
from the calibration lamp and subtracts that from the order flux.  Possible
line positions are identified by taking the derivative of the flux
versus pixel position with the IDL procedure \emph{DERIV} and looking for 
the positive to negative zero crossing that the positive and negative slopes
of an emission line create in the derivative.
A legitimate zero crossing requires two positive derivative values followed
by two negative derivative values.  The procedure does several 
checks on the legitimacy of the position.  The flux at the position must
be greater than a preset threshold which is equal to the median previously
subtracted from order flux.  If the line flux is only positive for one
pixel or if the line appears on the side of another line it is rejected.

After identification via the derivative method a Gaussian fit is done
at the positions that survived the checks described above to find the
line position with fractional pixel accuracy.  This is a straight Gaussian 
fit with no continuum as the continuum was subtracted earlier.  This is
done with a slightly modified version of the IDL procedure \emph{GAUSSFIT}
called \emph{gaussfitstat} that returns the status flag of the fitting 
procedure as well as the parameters of the Gaussian fit. Any position with 
a bad status flag, indicating an internal IDL error such as underflow or
overflow, is rejected.  The position
found by the Gaussian fit is placed in a new line structure along with the 
coefficients of the Gaussian fit.  Unlike the Th/Ar line list the new structure 
array only contains structures for the lines found in the median long-slit 
calibration line image. The structure for each line has 22 different entries 
which eventually include the information from the 
Th/Ar line lists, the ``Murphy'' flag, the spectral order and other information
for record keeping. One of the entries is a ``use'' flag on whether it
should be used in the final wavelength solution.  Not all lines with
positive ``Murphy'' flags were used in the solution.

After all of the line positions have been determined a preliminary assignment
of line identifications is made.  The preliminary wavelengths are matched
against an atomic line list described above in \S~\ref{sss-ta}. The line 
in the list with the closest wavelength to the measured wavelength of the line 
is assigned as the preliminary identification.  There are 3751 identified 
lines for the 390 grating setting and 3087 lines for the 430 setting. Most
of these lines are not be used in the wavelength calibration

\subsubsection{Using vacuum wavelengths} \label{sss-vw}

The wavelength calibration is performed using vacuum wavelengths.  
This means that the wavelength solution produced assigns the proper 
vacuum wavelength for each Th/Ar line as the appropriate wavelength
for that pixel position in each order.  The observations, however,
were in conditions nearer to STP
than vacuum.  The wavelength calibration solution simply gives the
vacuum wavelength as a function of pixel position.  As such it is
only valid for observations carried out under similar atmospheric
conditions to those present when the calibration observations were
taken.  This would be true whether we used STP wavelengths, current
atmosphere wavelengths or vacuum wavelengths.  The \citet{lov07}
measurements were made in vacuum therefore we have chosen vacuum 
wavelengths as our standard.  The shifts in temperature for any
night of observation were 0.1 degrees Centigrade or less and the
pressure shifts were 0.7 mm. Hg or less during a night as shown in 
Table~\ref{tab-lag}. At 4000 \AA\  the rate of change of air
wavelength is less than $4 \times 10^{-3}$ \AA\ per degree 
Centigrade and slightly more than $1 \times 10^{-5}$ \AA\ per
mm. Hg.  These are changes on the order of 1 part in $10^7$ or
less for all nights and are 1/3 or less of the accuracy of our 
wavelength solution. Since shifts are calculated on a night to
night basis and on an order by order basis, any shifts between
nights are accounted for in the zero point calculation for the
wavelengths of each order for a night.  Future improvements, such
as higher resolution spectra, will require a temperature and pressure
correction in the procedure.

\subsubsection{Refining the wavelength solutions}
\label{ss-rw}

The final step in the first pass through the wavelength calibration
is a refinement of the wavelength solution.  This is accomplished
with an interactive IDL procedure called \emph{waverefine\_uves} in 
Figure~\ref{fig-ov}.  The procedure provides a method of improving
the preliminary 6 term Legendre polynomial wavelength solution.
Wavelength solutions employing different polynomials and Legendre 
polynomial fits of orders 4 through 7 were tried initially.  The Legendre 
polynomial fits appeared to be the most stable, without high deviations
from acceptable fits at the end points of the solutions.  The quality
of fits improved up to the 6 order solution used in this work.  The 
addition of an additional term to make a 7 order fit did not improve
the fits and in some cases made them worse.  This choice is probably
unique to the UVES blue channel and may be different for other
spectrometers with different sources of image distortion.

For each order the procedure initially deems all identified lines as
not usable.  As each section of the spectrum for an order is displayed
on the monitor the user indicates which  lines should be included in
the wavelength solution.  A vertical line for each identified line
is displayed at the wavelength listed in the line list with a height 
that is proportional to the strength in the line list.  This helps the user
match the observed spectrum to the appropriate lines in the list.
If the line is one of the vetted acceptable lines identified in
\citet{mur07} then a letter M is printed above the line position.
During this first pass the majority of the lines are rejected based
on a poor signal-to-noise ratio, asymmetry due to blending or obvious double
lines.  Each order contains between 80 and 100 lines from the list
described above.  Usually about 25 lines are selected as usable.
The choices of usable lines are subjective but most of the rejections
are fairly obvious.  This rejection ratio is similar to the ratio
described in \citet{mur07}.  Our selection of lines
is independent of the \citet{mur07} selections in the following manner. 
Our inital selection and rejection of lines was done without reference
to which lines were Murphy lines. This led to the result that some lines that were
not Murphy lines are included and several Murphy lines were rejected.
For lines that are bluer than the \citet{lov07} list only the Murphy 
lines were available but many of those were rejected.  The statistics for
the 390 grating setting are that there are 2209 Murphy lines in the
spectral region of the observations.  Of these we chose 518 to use in
our analysis plus 2 that are not Murphy lines.  For the 430 grating 
setting there are 1462 Murphy lines in the spectral region of which
we chose 527 plus again 2 lines that are not Murphy lines. For obvious 
reasons most of the selected lines are Murphy lines due to similar 
selection criteria.  Almost all of the rejections of Murphy lines
in our list are due to a more conservative cut on line crowding. This
is a key difference in using the 2x2 binned pixels rather than the
individual CCD pixels.   Another difference is that our selection is
order by order for each of the grating settings.  A line that is 
selected in one order may be rejected in an adjacent order if it falls
in a poor signal to noise region.

After the first pass through the order the procedure produces the Legendre 
polynomial fit to the lines marked usable and displays the deviation
of the measured lines from the fit.  Typically there will be a few 
extreme outliers.  The program allows the user to add or remove lines from
the usable list one at a time.  The new solution is displayed after
each addition or removal.  This is the most subjective part of the procedure
and the most scientifically dubious.  The rejection of outlying
data points is risky.  The assumption here is that the true
wavelength solution should be smoothly varying without any 
discontinuities or high-frequency variation.  Practice with the
procedure leads to better identification of the outliers that
are causing the errors in the solution.  Often it is not the
furthest outlier that is causing a problem.  Removal of a single 
outlier that produces discontinuous break between positive and
negative outliers often greatly reduces the average error.

Another subjective aspect to using the procedure is that the
final set of deviations from the solution should be ``random''.
A solution that produces a sharp break in the sign of the outliers
or that has a systematic trend in the location of the outliers is
probably wrong.  After the solutions are adjusted typically 20 of
the original 25 lines are left as usable.  After all of the
orders have their solutions the procedure is repeated several
times to improve the fit.  This allows some lines that
may have been erroneously rejected in a previous pass to be 
returned to the usable line list and may result in rejection
of other lines.

\subsubsection{First pass data products}

The first pass data products are: the first pass wavelength solutions
which is a set of the 6 coefficients of the first 6 Legendre polynomials
for each order that represent the wavelength solution for that order
in terms of \AA\ versus pixel;  a median long-slit calibration image, 
the first pass one dimensional order by order spectrum;  and the long 
slit calibration line structure array with the use flag set positive
for all of the lines that were deemed usable in the first pass.  All of 
these products are 
updated in the second pass through the wavelength solution.  The flat
correction image, order definition solutions and the length of the slit
are first pass data products that are not updated in the second pass.

\subsection{Second pass wavelength solution}

\begin{figure}
\includegraphics[scale=.6]{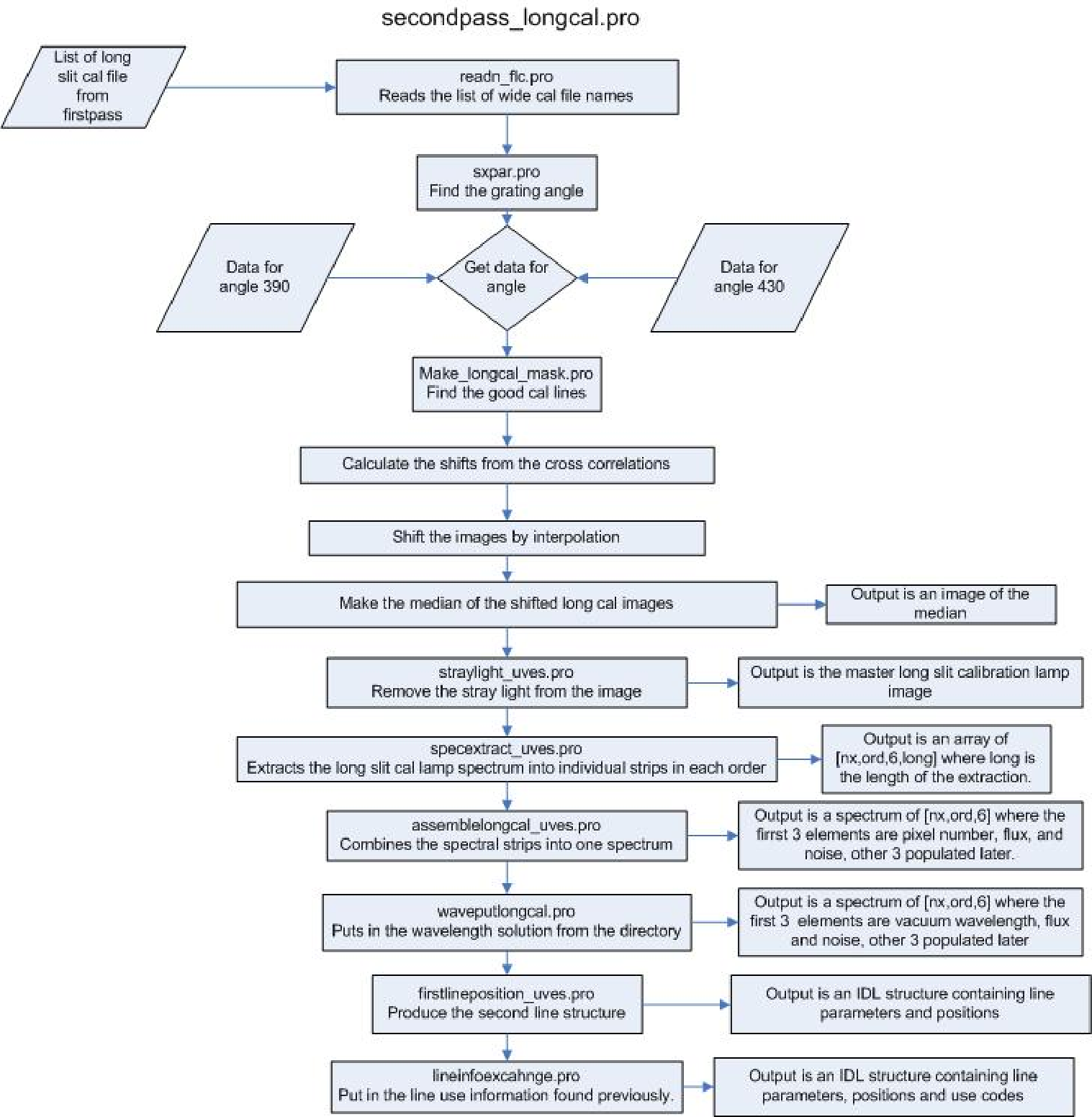}
\caption{Flow chart for the second pass analysis of the long-slit calibration
lamp spectrum and wavelength calibration. The names that end in .pro at the
top of the boxes are the names of the IDL based procedures that we developed.}
\label{fig-sp}
\end{figure}

The second pass to the wavelength calibration recognizes that the
individual long-slit calibration lamp images taken over the course
of the observations are not perfectly aligned.  It computes the 
offsets between the images using the lines that were declared 
usable in the first pass.  The procedure shifts the images to a common
registration and produces a new median image which becomes the 
master long-slit calibration lamp image.  It then follows most of
the steps used in the first pass to produce a new wavelength solution
for the master long-slit calibration lamp image.  It also preserves
the shifts calculated for the individual images to provide a baseline
to compute the shifts appropriate for the source spectrum images.
A flow chart of the second pass procedures is given in Figure~\ref{fig-sp}.

\subsubsection{Calculating the long-slit calibration image offsets}

The procedure starts by reading in the list of long-slit calibration
lamp images and the line structure array generated in the first pass.
For each line marked usable the procedure \emph{make\_longcal\_mask} finds
the corner coordinates in pixels of a 7 by 21 rectangle in x and y
centered on the line position.  These small sub-images are the regions
that will be cross correlated to find the shifts between the images.
The first long-slit calibration lamp image is arbitrarily chosen as
the reference image.  For each line in each image the small two dimensional
image is summed in the y cross dispersion direction to form a 7 pixel 
one dimensional spectrum.  The one dimensional spectra are interpolated 
to a grid that has a 0.01 pixel spacing and cross correlated with the 
corresponding interpolated spectrum in the reference spectrum.  The 
lag producing the maximum cross
correlation signal is selected as the shift for that line relative
to the reference image.  After all of the shifts have been calculated
the shift for each long-slit calibration line image is taken as the 
median of the shifts for all of the usable lines in the image.  Histograms
of the shifts for each image are also plotted to check for anomalous
distributions of shifts.  The shifts for the 390 and 430 images are
given is Table~\ref{tab-lag}.

Once the shifts have been calculated each image is shifted to the 
position of the zero shift image by interpolation.  The IDL procedure 
\emph{INTERPOLATE} is used with the value of the CUBIC parameter set to
-0.5 and the keyword GRID set to true.  The value of CUBIC is the 
cubic polynomial interpolation parameter.  It is the value of $\alpha$
in the following cubic polynomials r(x) which are used to calculate the
contribution to the interpolated value from adjacent pixels, see 
\citet{par83}.

$ r(x) = (\alpha + 2) \mid x \mid^3 - (\alpha+3) \mid x \mid^2 +1 $
\hspace{.5in} $\mid x \mid < 1$

$ r(x) = \alpha \mid x \mid^3 - 5 \alpha \mid x \mid^2 + 8 \alpha
\mid x \mid - 4 \alpha$ \hspace{.5in} $1 \leq \mid x \mid \leq 2$

$ r(x) = 0$ otherwise

\noindent The shifted images are median combined to produce a 
cosmic ray free long-slit calibration lamp image.  The stray light 
removal procedure described earlier in
\S~\ref{ss-sl} is performed on the median image and the result is 
stored as the master long-slit calibration lamp image.  Future
long-slit calibration lamp images can be added to the master image
by repeating the processing with the full set of images.

The second pass spectrum extraction and spectrum assembly are identical
to the first pass processing.  The wavelength solution found from the 
first pass is put into the spectrum. 
The IDL procedure \emph{firstlineposition.pro} again assembles the detected
lines into a line structure array with the new parameters from the
master long calibration lamp spectrum.  Rather than starting from
scratch the line usage flags from the first pass are transferred 
from the first pass line structure array to the second pass array
by the procedure \emph{lineinfoexchange.pro}.

The wavelength solution is again refined
in iterative passes as described in \S~\ref{ss-rw}.  The result
of this procedure is the master wavelength solution.  Out of the 3751
identified lines in the 390 grating setting 686 lines were used
in the wavelength solution. Only 4 are not in the vetted Murphy
list.  577 lines out of the 3087 identified lines were used in the 
430 grating setting wavelength solution, 3 of which were not in the
Murphy list.  The difference between these numbers and the numbers
in \S~\ref{ss-rw} is that here a line is counted more than once if
it appears in more than one order.  The use rate of the identified 
lines iss $18\%$ for the 390 setting and $19\%$ for the 430 setting.  
This is lower than the retention fraction in the \citet{mur07} list.

\subsubsection{Final wavelength calibration data products}

The final data products for each grating setting are stored for use
in determining the wavelength solution for the object spectra.  They
include the master long-slit calibration lamp image, the master one
dimensional order by order spectrum, the master wavelength solution for 
that spectrum and the associated structure array of line parameters.  
Also stored are the wavelength shifts in pixels for each 
of the individual long-slit calibration lamp images and the times that 
they were taken.  The last data sets are the median flat correction image, 
the order definition solutions and the length of the long-slit in pixels
from the first pass solutions.  A list of the UVES Thorium and
Argon lines used for the 390 and 430 grating settings wavelength solutions
is included in the electronic version of this paper.  These lists are
particular to the grating settings as opposed to the vetted line lists
in \citet{mur07}.  This is because a line that has a good signal-to-noise ratio
in one order may fall in a poor signal-to-noise ratio area in an adjacent order.

\subsection{The form of the wavelength solutions}

The wavelength solutions are given by

\begin{equation}
\lambda_i = \sum_{i=0}^{5} C_i L_i(x)
\end{equation}

where $L_i$ are the Legendre polynomials and the $C_i$ are the coefficients
for each order that are the products of \emph{refinewave\_uves.pro}.  The Legendre
polynomials are only valid in the range between -1 and 1 therefore the
variable x is related to the pixel number p by 

\begin{equation}
x = \frac{p-750.}{750.}
\end{equation}

where p ranges between 0 and 1499 for the 1500 pixel width of the blue channel.
This means that the largest value of x is 0.99866666 rather than 1 but that
is of no consequence since each value of x is unique, equally spaced and 
within the valid range of the Legendre polynomials.  Given the nature of the
Legendre polynomials the coefficient $C_0$ is the wavelength in \AA\
of the middle of the order and $C_1$ is the linear dispersion in \AA\
per value of x.  The higher order coefficients describe the deviation from
linearity of the wavelength solution.  Tables \ref{tab-390} and \ref{tab-430}
give the coefficients found for the master long-slit spectra at the two
grating settings.  Since all of the long-slit calibration images were shifted
to the position of the first calibration image for each run the solutions
presented here represent the best wavelength solution for the first
calibration image.  Since the shifts are small between images these solutions
can be used as starting points for other spectra at those grating settings.

\subsection{Accuracy of the solutions} \label{ss-as}

We can estimate the accuracy of the wavelength solution by looking at 
how accurately the 6 term Legendre polynomial solution reproduces the
measured vacuum wavelengths of the calibration lines used in producing
the solution.  The statistical look at the fits is appropriate since
for most orders the number of lines used in the solution is 3 to 4
times the number polynomial terms.  The statistics of the fits for
all of the lines gives a standard deviation on the fractional error
$\Delta\lambda/\lambda$ of $2.6 \times 10^{-7}$ for the 390 grating
setting and $2.2 \times 10^{-7}$ for the 430 grating setting.  A
listing of the fractional errors for each order used
in the program described by \citet{thm08} is
given in table \ref{tab-stat}.

\begin{figure}
\includegraphics[scale=0.6]{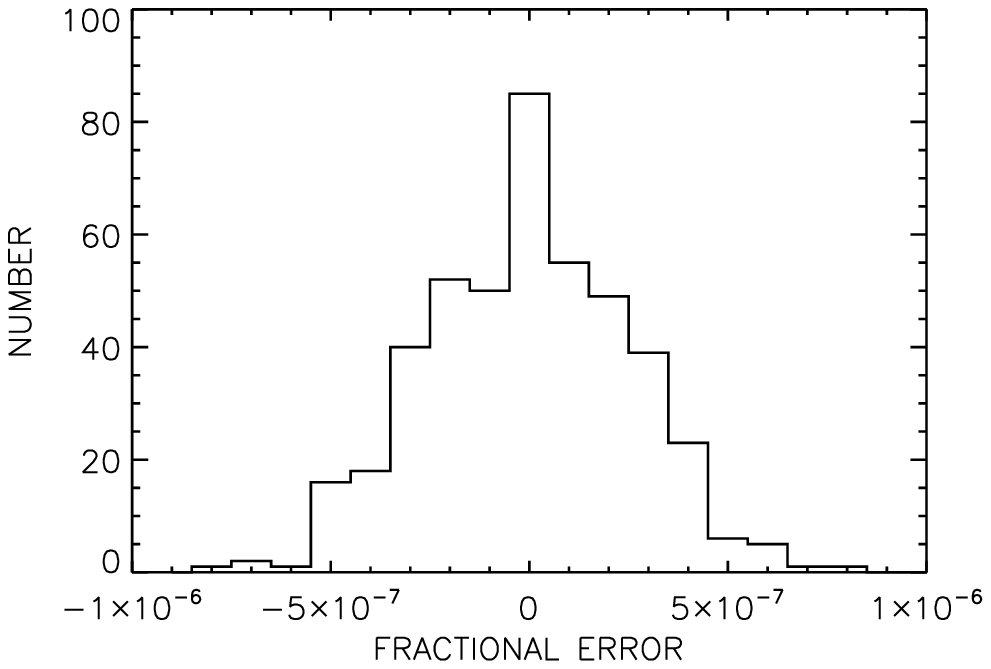}
\caption{The histogram of the line wavelength fitting fractional
residuals $\Delta\lambda / \lambda$ for the 390 grating setting}
\label{fig-wh3}
\end{figure}

\begin{figure}
\includegraphics[scale=0.6]{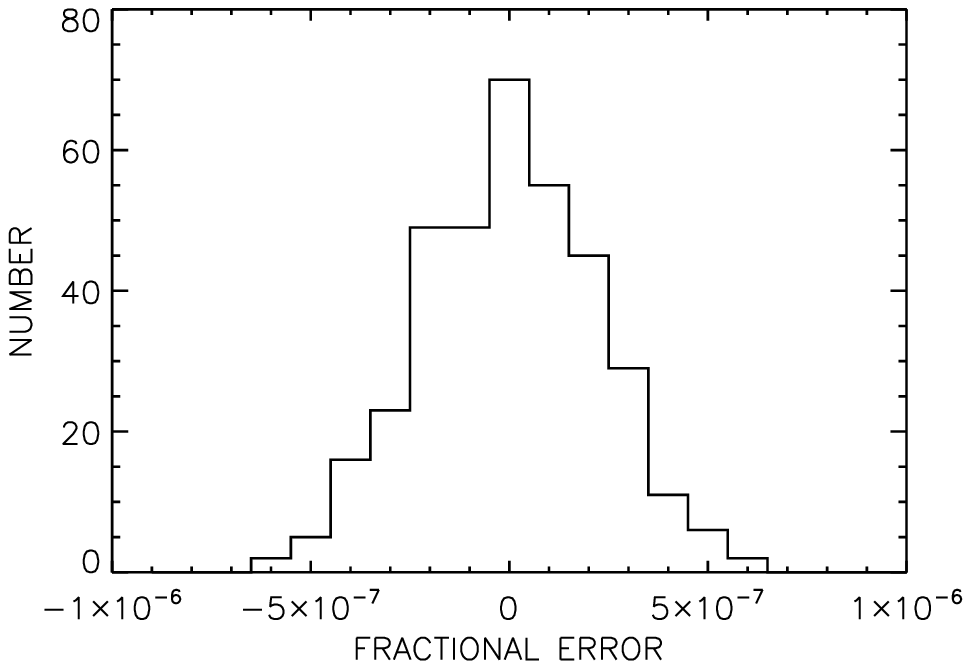}
\caption{The histogram of the line wavelength fitting fractional
residuals $\Delta\lambda / \lambda$ for the 430 grating setting}
\label{fig-wh4}
\end{figure}

The distribution of errors is shown in Figures~\ref{fig-wh3} and
\ref{fig-wh4}
where the fractional differences between the calculated and 
listed wavelengths are shown in histograms with a bin size
of $1 \times 10^{-7}$.  The histograms are reasonably 
symmetric about zero and the $1\sigma$ values correspond to
roughly 75 m s$^{-1}$.  Comparison with Figure 8 in 
\citet{mur07} indicates that the wavelength errors are a
factor of 2 less than the pipeline values but somewhat more
than achieved by \citet{mur07} for their data sets which contain
more calibration line images than this work, a narrower slit
width (0.6 rather than 0.8 arc seconds) and unbinned images.
In both cases the line selection is the primary component in
improving the accuracy of the calibration. The data sets
we use, however, are the calibration line spectra taken at the
time of the observations we want to calibrate and better 
represent the wavelength calibration for those objects.  

\begin{figure}
\includegraphics[scale=0.6]{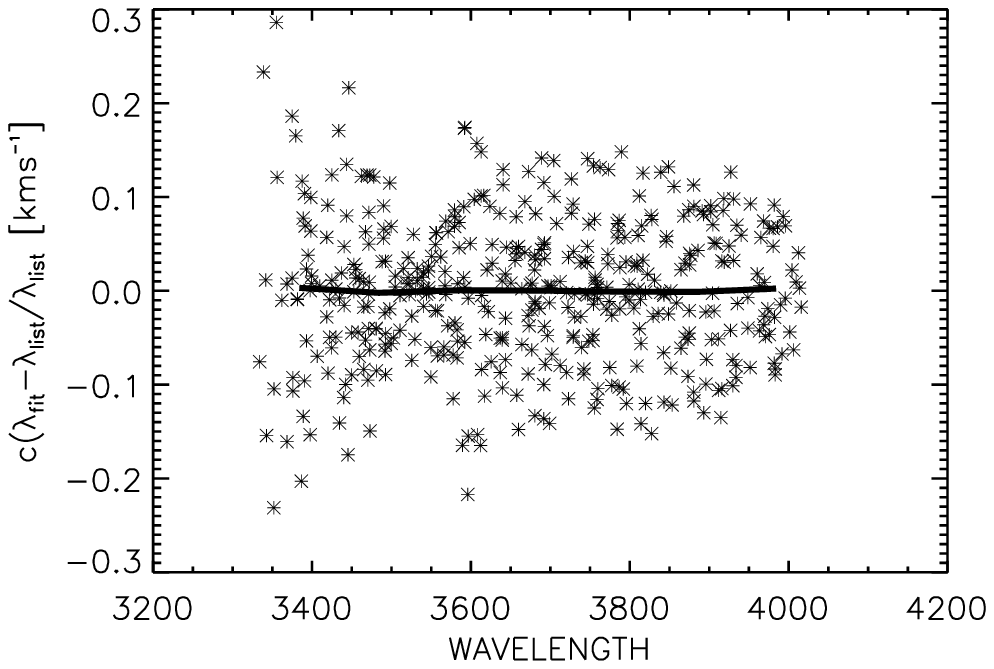}
\caption{The plot of all of the wavelength fitting fractional
residuals $\Delta\lambda / \lambda$ for the 390 grating setting.  The
solid black line is the mean of the residuals in 100 \AA\ bins
to check for systematic residuals.}
\label{fig-ep3}
\end{figure}

\begin{figure}
\includegraphics[scale=0.6]{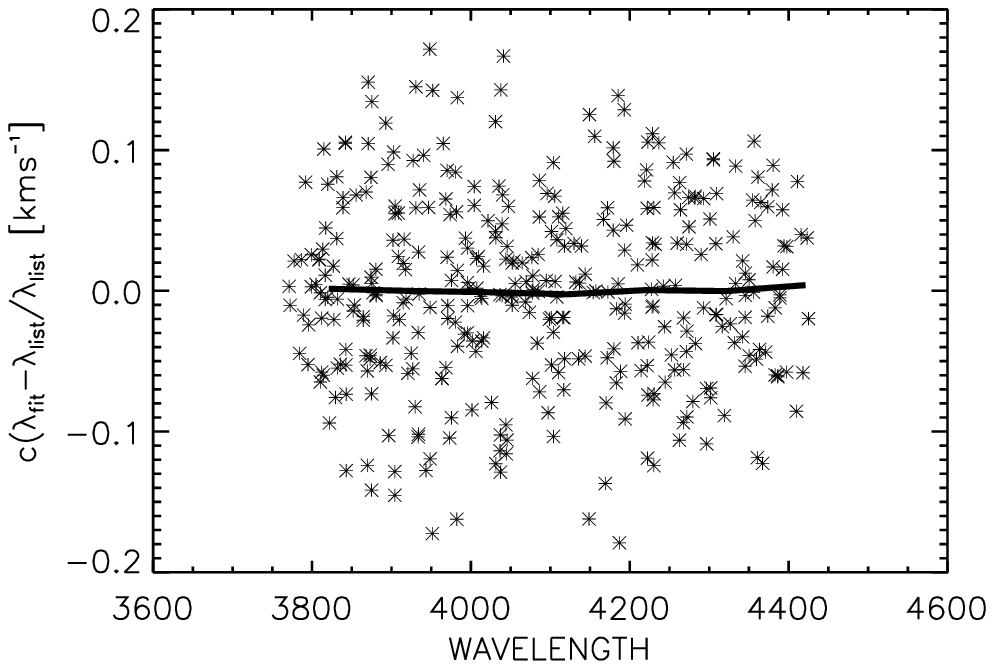}
\caption{The plot of all of the wavelength fitting fractional
residuals $\Delta\lambda / \lambda$ for the 430 grating setting.  The
solid black line is the mean of the residuals in 100 \AA\ bins
to check for systematic residuals.}
\label{fig-ep4}
\end{figure}

One of the primary points of \citet{mur07} is that the pipeline
UVES wavelength calibration has significant systematic residuals
as shown in their Figure 9 and that this residual can give false
indications of variation in the measurement of the fundamental
constant $\alpha$.  The same systematic error residual can 
affect the measurement of $\mu$ as well but in a different 
manner \citep{thm08} than for $\alpha$.  Since the primary purpose
of this paper is to document the wavelength calibration for
the determination of $\mu$ at early times in the universe we
have carried out an analysis similar to that of \citet{mur07}
for our wavelength calibration.  Figures~\ref{fig-ep3} and
\ref{fig-ep4} show the
scatter of all of the calibration line fits that are used in the
determination of $\mu$ at both the 390 and 430 grating settings.
This figure can be directly compared to Figure 9 in \citet{mur07}.
The solid black line is the mean residual in 100 \AA\ bins.  It
is clear that the large scale residuals present in the UVES pipeline
calibration are not present in the wavelength calibration presented
here.  The minimum near 3550 \AA\ in figure~\ref{fig-ep3} is not
well understood but may be simply a statistical anomaly.

\section{Addition and wavelength solution of the individual object spectra}

Once the wavelength calibration has been performed it can be applied
to the individual spectra of the objects produced by the pipeline.
A typical script for the pipeline reduction is given in 
Appendix~\ref{a-red}  The standard output of
the UVES pipeline are spectra that have been interpolated onto a
constant increment wavelength grid.  All orders have been combined
to produce a continuous spectrum with no order breaks.  The details
of this process are not immediately obvious from the pipeline
documentation \citep{bal04}.  The results of the
optimal extraction described in \citet{bal04} are, however, available
as an intermediate product.  These files for the blue channel
are labeled fxb\_UVES*\_b.fits, where the * represents a date and
time tag along with a 3 digit identifier.  These are order by order
one dimensional optimal extractions of the spectra.
  
\subsection{The wavelength solutions for the object spectra}

The object spectra wavelength solutions start with a determination
of the shift in wavelength
of the object spectrum relative to the master wavelength solution.
This is done by fitting the shifts of the individual long-slit
calibration lamp images versus time for each night of observing.
For nights with just two images this is a linear fit.  Nights
with 3 images are fit with a quadratic fit.  The shifts for the
object spectra are determined from the shift versus time
fit function by the midpoint in time of the integration.  The shifts
are similar to the shifts for the calibration spectra shown in
Table~\ref{tab-lag} which are on the order 0.01-0.02 pixels. The
master solution is interpolated to the shifted positions found
above and placed into the wavelength section of the spectrum array.  This
wavelength solution is the appropriate vacuum wavelengths for a
source at rest relative to the spectrometer and therefore has to
be corrected for barycentric velocity.  The appropriate barycentric
velocity is calculated for the midpoint of the observation and
the wavelength solution is then  corrected to account for the 
barycentric velocity of the earth at the location of the VLT. 
At this point in the analysis the center points of the pixels in 
the different object spectra have different wavelength
solutions.

In preparation for addition of the spectra the spectra are all
shifted to the wavelength solution of the master wavelength solution
by interpolation.  The interpolation shift is determined from the
shift relative to the master solution calculated above.  This is not
done initially since the various spectra have slightly different
barycentric velocity contributions.  As with the long-slit calibration
spectra that bracket the observations, the shifts from the master solution
is generally on the order of a few hundredths of a pixel.  The 
spectra now all have the same pixel wavelength solution and can 
be added together.

\subsection{Addition of the spectra}

Since the spectra now all have the same wavelength per pixel solution 
they can be directly added together for each order.  The 9 spectra for
each object are combined to produce both the median and the mean for each
pixel.  The median better rejects any residual cosmic ray hits in the
spectra and provides the primary spectrum for the analysis of the value
of $\mu$ in \citet{thm08}.  The mean spectrum is also analyzed to provide a 
check on the median spectrum. Other ``optimal'' additions methods were
utilized that weight the pixel flux by its signal to noise ratio.
These methods did not produce quantifibly better results than the 
median spectra.  This is most probably due to the remarkable uniformity
of the spectra between observations of the same object.  For this reason 
we utilize the median spectra in our analysis.  Data sets with significantly
varying signal to noise ratios between spectra will certainly benefit from
more optimal spectrum addition methods. 

\section{Conclusions}

The measurement of fundamental constants through astronomical
spectra requires the best possible wavelength calibration that
can be achieved.  This requires both observing and data reduction
procedures beyond what is practical for most standard observation
and data reduction procedures.  The procedures described in this
paper have achieved a $1\sigma$ fractional $\Delta \lambda / 
\lambda$ wavelength accuracy of better than $3 \times 10^{-7}$
for UVES spectra which is a factor of 2 better than the accuracy
of the UVES pipeline reduction.  More importantly there is no
detectable systematic residual wander as was discovered by
\citet{mur07} in the pipeline wavelength calibration.

The reduction methods are described in detail so that they can
be repeated by other researchers and a suggested observing
sequence was presented that can improve the accuracy of both
the wavelength calibration and its transfer to the object
spectra.  This paper serves as a detailed description of
the calibration procedures used in the determination of the 
fundamental constant $\mu$ through the molecular hydrogen
absorption lines in the spectra of two QSOs \citet{thm08}. 

\section{Acknowledgments}
RIT wishes to acknowledge and thank Wim Ubachs for several
discussions and comments that have greatly improved this paper.
RIT also wishes to acknowledge the helpful comments of an 
anonymous referee that greatly benefited this paper. C.M wishes
to acknowledge very useful discussion with Paolo Molaro.
The work of C.M. is funded by a Ci\^encia2007 Research Contract.



\clearpage

\begin{table}[h]
\begin{tabular}[t]{cccccccc}
\hline
\hline
\multicolumn{4}{c}{Grating Setting} & \multicolumn{4}{c}{Grating Setting}\\
\hline
\multicolumn{4}{c}{390} & \multicolumn{4}{c}{430} \\
\hline
MJD Date & shift & Temp. & Press & MJD Date & shift & Temp. & Press  \\
JD-2.4E6 & pixels & C$^o$ & mm. Hg & JD-2.4E6 & pixels & C$^o$ & mm. Hg \\
\hline
\hline
52643.08644 & 0.00 & 14.8 & 743.6 & 52282.08493 & 0.00 & 11.8 & 745.6 \\
52643.14617 & -0.06 & 14.8 & 743.6 & 52282.13882 & -0.02 & 11.8 & 745.7 \\
52643.20522 & -0.05 & 14.7 & 743.0 & - & - & - & - \\
\hline
52644.09176 & -0.02 & 14.5 & 743.3 & 52283.08329 & 0.03 & 11.2 & 746.4 \\
52644.21381 & -0.03 & 14.4 & 742.2 & 52283.13777 & 0.04 & 11.2 & 746.4 \\
\hline
52645.09255 & 0.00 & 13.5 & 744.2 & 52284.08694 & 0.01 & 11.5 & 745.6 \\
52645.15520 & 0.01 & 13.5 & 743.9 & 52284.14110 & 0.02 & 11.5 & 745.7 \\
52645.21402 & -0.02 & 13.5 & 743.2 & - & - & - & - \\
\hline
\end{tabular}
\caption{Dates, shifts in position in pixels, temperature and pressure for the
390 and 430 grating setting long-slit calibration lamp
images}
\label{tab-lag}
\end{table}
\clearpage

\begin{table}
{\scriptsize
\begin{tabular}[t]{ccccccc}
\hline
order & C$_0$ & C$_1$ & C$_2$ & C$_3$ & C$_4$ & C$_5$ \\
139 &  3.3578451E+03 &  2.6241278E+01 & -2.2600063E+00 &  4.0406710E-02 & -1.5302832E-02 &  8.2013765E-03 \\
138 &  3.3821772E+03 &  2.6446218E+01 & -2.2865587E+00 &  6.7419982E-02 & -1.3854520E-02 &  1.6782558E-02 \\
137 &  3.4068802E+03 &  2.6582786E+01 & -2.2671584E+00 &  9.5942609E-03 &  6.7850203E-03 & -9.2851399E-03 \\
136 &  3.4319311E+03 &  2.6778913E+01 & -2.3022740E+00 &  2.7985127E-02 & -8.5313181E-03 & -2.0850798E-03 \\
135 &  3.4573574E+03 &  2.6966587E+01 & -2.3222197E+00 &  3.5873646E-02 & -1.2983645E-02 &  1.2479542E-03 \\
134 &  3.4831663E+03 &  2.7155877E+01 & -2.3347319E+00 &  3.7873325E-02 & -6.2703785E-03 &  1.1789081E-03 \\
133 &  3.5093650E+03 &  2.7334353E+01 & -2.3352488E+00 &  1.9924731E-02 &  3.2293852E-03 & -7.4094784E-03 \\
132 &  3.5359529E+03 &  2.7538598E+01 & -2.3656308E+00 &  3.3747420E-02 & -9.2136909E-03 & -1.5741166E-03 \\
131 &  3.5629531E+03 &  2.7732403E+01 & -2.3769408E+00 &  2.9649622E-02 & -7.1513702E-03 & -1.9546620E-03 \\
130 &  3.5903666E+03 &  2.7933761E+01 & -2.3960270E+00 &  3.4519896E-02 & -7.7112736E-03 &  3.4520805E-04 \\
129 &  3.6182069E+03 &  2.8136593E+01 & -2.4112124E+00 &  3.4322472E-02 & -1.3197668E-02 &  3.3426049E-04 \\
128 &  3.6464797E+03 &  2.8344177E+01 & -2.4316644E+00 &  3.6990870E-02 & -1.1417702E-02 &  5.6447434E-04 \\
127 &  3.6751965E+03 &  2.8560555E+01 & -2.4575814E+00 &  4.3546365E-02 & -1.7266535E-02 &  6.9978172E-03 \\
126 &  3.7043758E+03 &  2.8764023E+01 & -2.4596247E+00 &  3.2936489E-02 & -7.9666039E-03 & -7.1235356E-04 \\
125 &  3.7340186E+03 &  2.8980703E+01 & -2.4752705E+00 &  3.0423652E-02 & -8.3680162E-03 & -1.1760229E-03 \\
124 &  3.7641382E+03 &  2.9204603E+01 & -2.4968770E+00 &  3.4123134E-02 & -1.0928570E-02 & -7.7307545E-04 \\
123 &  3.7947486E+03 &  2.9427401E+01 & -2.5117336E+00 &  3.2840418E-02 & -7.2320179E-03 & -3.6479544E-03 \\
122 &  3.8258608E+03 &  2.9656143E+01 & -2.5307968E+00 &  3.1801252E-02 & -6.5872613E-03 & -2.3128226E-03 \\
121 &  3.8574873E+03 &  2.9886083E+01 & -2.5518506E+00 &  2.9777719E-02 & -8.6517922E-03 & -6.3210983E-03 \\
120 &  3.8896425E+03 &  3.0121318E+01 & -2.5652903E+00 &  2.9130129E-02 & -2.0933691E-03 & -6.0873269E-03 \\
119 &  3.9223364E+03 &  3.0368623E+01 & -2.5902555E+00 &  3.4140415E-02 & -1.3312468E-02 & -2.9620043E-03 \\
118 &  3.9555834E+03 &  3.0611776E+01 & -2.6114255E+00 &  3.6610362E-02 & -9.5989623E-03 & -1.7570264E-03 \\
117 &  3.9894010E+03 &  3.0858347E+01 & -2.6269035E+00 &  3.0456590E-02 & -3.8195133E-03 & -4.5113309E-03 \\
116 &  4.0237990E+03 &  3.1117735E+01 & -2.6542921E+00 &  3.6451959E-02 & -7.5083039E-03 & -2.7498142E-03 \\
\end{tabular}
}
\caption{The coefficients of the first 6 Legendre polynomials C$_0$-C$_5$
in the long-slit calibration lamp image for the 390 grating setting are
listed in the table.}
\label{tab-390}
\end{table}
\clearpage

\begin{table}
{\scriptsize
\begin{tabular}[t]{ccccccc}
\hline
order & C$_0$ & C$_1$ & C$_2$ & C$_3$ & C$_4$ & C$_5$ \\
139 &  3.7640555E+03 &  2.9247633E+01 & -2.4983034E+00 &  2.1809631E-02 & -1.5302941E-03 & -7.3419534E-03 \\
138 &  3.7946624E+03 &  2.9476363E+01 & -2.5209971E+00 &  2.7012675E-02 & -5.1708948E-03 & -3.3158614E-03 \\
137 &  3.8257730E+03 &  2.9708029E+01 & -2.5427480E+00 &  3.1988803E-02 & -9.4145740E-03 & -1.3762068E-03 \\
136 &  3.8573989E+03 &  2.9940130E+01 & -2.5606036E+00 &  3.3863295E-02 & -9.1725422E-03 &  1.5817056E-03 \\
135 &  3.8895535E+03 &  3.0170043E+01 & -2.5755437E+00 &  2.6638458E-02 & -7.7280717E-03 & -6.9431765E-03 \\
134 &  3.9222446E+03 &  3.0413128E+01 & -2.5979035E+00 &  3.2158675E-02 & -9.7994044E-03 & -1.4903875E-03 \\
133 &  3.9554912E+03 &  3.0656432E+01 & -2.6178296E+00 &  3.3148307E-02 & -6.8278925E-03 & -1.5417298E-03 \\
132 &  3.9893084E+03 &  3.0903140E+01 & -2.6327016E+00 &  2.7640170E-02 & -5.7173851E-03 & -4.5849886E-03 \\
131 &  4.0237052E+03 &  3.1159181E+01 & -2.6591523E+00 &  3.4567971E-02 & -7.0897964E-03 & -2.5973397E-03 \\
130 &  4.0587017E+03 &  3.1416673E+01 & -2.6778756E+00 &  3.5027143E-02 & -9.9222133E-03 & -2.2467367E-03 \\
129 &  4.0943163E+03 &  3.1668129E+01 & -2.6874486E+00 &  2.3789638E-02 &  9.4800526E-04 & -8.3427076E-03 \\
128 &  4.1305553E+03 &  3.1948742E+01 & -2.7185913E+00 &  4.1422843E-02 & -8.0693347E-03 &  2.1152305E-03 \\
127 &  4.1674438E+03 &  3.2213865E+01 & -2.7383569E+00 &  2.9673746E-02 & -3.1165309E-03 & -7.7851206E-03 \\
126 &  4.2049961E+03 &  3.2492623E+01 & -2.7655304E+00 &  3.4471338E-02 & -8.6897107E-03 & -5.5739705E-03 \\
125 &  4.2432354E+03 &  3.2768542E+01 & -2.7788524E+00 &  2.7706485E-02 & -2.3526557E-03 & -5.2609012E-03 \\
124 &  4.2821701E+03 &  3.3066335E+01 & -2.8149856E+00 &  3.9617907E-02 & -1.1644463E-02 & -6.7238977E-04 \\
123 &  4.3218291E+03 &  3.3357925E+01 & -2.8360519E+00 &  3.8452345E-02 & -1.1566467E-02 & -1.5868857E-03 \\
122 &  4.3622303E+03 &  3.3658196E+01 & -2.8606765E+00 &  3.7344307E-02 & -7.4652479E-03 & -5.7107517E-03 \\
121 &  4.4033926E+03 &  3.3960171E+01 & -2.8889830E+00 &  3.1228066E-02 & -1.1341092E-02 & -7.0300975E-03 \\
120 &  4.4453403E+03 &  3.4275396E+01 & -2.9108898E+00 &  3.7639151E-02 & -1.1469007E-02 & -1.2729088E-03 \\
119 &  4.4880943E+03 &  3.4594699E+01 & -2.9380710E+00 &  4.0701831E-02 & -8.3199880E-03 &  1.7192203E-03 \\
118 &  4.5316788E+03 &  3.4918293E+01 & -2.9638680E+00 &  3.8092495E-02 & -1.1269793E-02 & -3.9247857E-03 \\
117 &  4.5761167E+03 &  3.5253461E+01 & -2.9941438E+00 &  4.5003528E-02 & -1.1020847E-02 &  3.4696559E-03 \\
116 &  4.6214372E+03 &  3.5589023E+01 & -3.0209908E+00 &  3.9108559E-02 & -6.1143803E-03 & -2.0235974E-03 \\
\end{tabular}
}
\caption{The coefficients of the first 6 Legendre polynomials C$_0$-C$_5$
in the long-slit calibration lamp image for the 430 grating setting are
listed in the table.}
\label{tab-430}
\end{table}
\clearpage

\begin{table}[h]
\begin{tabular}[t]{cccccccc}
\hline
\hline
\multicolumn{4}{c}{Grating Setting} & \multicolumn{4}{c}{Grating Setting}\\
\hline
\multicolumn{4}{c}{390} & \multicolumn{4}{c}{430} \\
\hline
Order & Std. Dev. & Largest Err. & Num. Lines & Order & Std. Dev. & Largest Error & Num. Lines \\
\hline
139& 5.40E-07& 9.54E-07& 13 & 123& 1.16E-07& 2.57E-07& 16\\
138& 3.72E-07&-6.77E-07& 14 & 122& 1.99E-07& 3.52E-07& 22\\
137& 2.08E-07&-3.20E-07& 14 & 121& 2.61E-07& 4.94E-07& 19\\
136& 3.62E-07& 7.22E-07& 17 & 120& 2.35E-07&-4.85E-07& 20\\
135& 2.67E-07&-4.98E-07& 21 & 119& 2.35E-07&-4.28E-07& 25\\
134& 1.98E-07& 4.12E-07& 20 & 118& 3.74E-07&-5.75E-07& 18\\
133& 1.65E-07& 3.83E-07& 15 & 117& 1.65E-07& 3.49E-07& 21\\
132& 1.06E-07&-2.47E-07& 16 & 116& 1.82E-07&-4.30E-07& 22\\
131& 1.83E-07&-3.06E-07& 21 & 115& 2.70E-07& 5.56E-07& 23\\
130& 3.92E-07&-7.24E-07& 24 & 114& 1.58E-07&-3.45E-07& 23\\
129& 2.38E-07& 3.77E-07& 19 & 113& 1.59E-07& 3.04E-07& 13\\
128& 2.59E-07&-4.93E-07& 15 & 112& 3.29E-07&-5.97E-07& 21\\
127& 1.96E-07& 4.24E-07& 25 & 111& 1.90E-07&-3.97E-07& 23\\
126& 2.78E-07& 4.72E-07& 24 & 110& 2.44E-07&-4.14E-07& 21\\
125& 2.15E-07& 4.70E-07& 20 & 109& 1.96E-07&-3.13E-07& 20\\
124& 2.64E-07& 4.44E-07& 25 & 108& 1.99E-07&-3.63E-07& 17\\
123& 2.37E-07& 4.94E-07& 25 & 107& 2.12E-07&-4.09E-07& 22\\
122& 2.81E-07&-5.08E-07& 19 & 106& 1.76E-07&-2.86E-07& 16\\
121& 2.42E-07& 4.41E-07& 19 & - & - & - & - \\
120& 2.56E-07&-4.33E-07& 21 & - & - & - & - \\
119& 2.58E-07&-4.51E-07& 19 & - & - & - & - \\
118& 2.26E-07& 3.26E-07& 18 & - & - & - & - \\
117& 1.72E-07&-2.78E-07& 21 & - & - & - & - \\
\end{tabular}
\caption{Statistics for the fractional errors in wavelength
for the 390 and 430 grating setting wavelength solutions.}
\label{tab-stat}
\end{table}
\clearpage

\clearpage

\clearpage

\clearpage

\clearpage

\clearpage

\appendix

\section{A typical UVES pipeline reduction script used in this work}
\label{a-red}

The following is a set of typical commands is used with the MIDAS based
UVES pipeline to produce the pipeline products used in this analysis.
It is presented for completeness and as an example of the process.  
Individual batch reductions may vary slightly from the example.  The
commands are issued in the directory that contains the data recovered
from the archive. This script was for the Q0347-383 spectra at the 
430 grating setting. 

flmidas

confit/displ

split/uves refcalb.cat refcalb\_split.cat

create/icat referb.cat null do\_classification

add/icat referb.cat thargood\_2.tbl

prepare/caldb refcalb\_split.cat referb.cat

split/uves object.cat object\_split.cat

reduce/uves object\_split.cat object\_redb.cat ref430.2x2.cat E optimal median

After the script was run the appropriate bdf files were converted to FITS
format using the outdisk/fits command.

\end{document}